1

# Xtoys: cellular automata on xwindows

Michael Creutz[a] [*]

[a]Physics Department, Brookhaven National Laboratory, PO Box 5000, Upton, NY 11973-5000, USA
creutz@wind.phy.bnl.gov

Xtoys is a collection of xwindow programs for demonstrating simulations of various statistical models. Included are xising, for the two dimensional Ising model, xpotts, for the $q$-state Potts model, xautomalab, for a fairly general class of totalistic cellular automata, xsand, for the Bak-Tang-Wiesenfeld model of self organized criticality, and xfires, a simple forest fire simulation. The programs should compile on any machine supporting xwindows.

## 1. INTRODUCTION

The URL "http://penguin.phy.bnl.gov/www/xtoys/xtoys.html" on the World Wide Web points to a set of cellular automata simulators for Xwindows. Included are: xising, an Ising model simulator, xpotts, for the Potts model, xautomalab, for cellular automata, xsand, an sandpile model, and xfires, a simple forest fire automaton.

To run these on your workstation, get the file "xtoys.tar.Z," do "uncompress xtoys.tar.Z", then "tar -xvf xtoys.tar" and finally "make". The programs are freely distributable.

These should all compile under generic X; if something does not work on your machine supporting X, let me know so I can fix it. If you encounter difficulties compiling, possibly the X11 include files are not being found. Then you need to compile with a -I option to where they are and possibly change the -lX11 to help the linker. As with any Xwindow program, you need to have "xhost" and "DISPLAY" set up properly.

The remaining sections summarize some experiments with these programs. You may wish to run them concurrently with reading the remainder of this document. The xautomalab display is shown in Fig. (1).

<>
[*]This manuscript has been authored under contract number DE-AC02-76CH00016 with the U.S. Department of Energy. Accordingly, the U.S. Government retains a non-exclusive, royalty-free license to publish or reproduce the published form of this contribution, or allow others to do so, for U.S. Government purposes. Report no. BNL-62123.

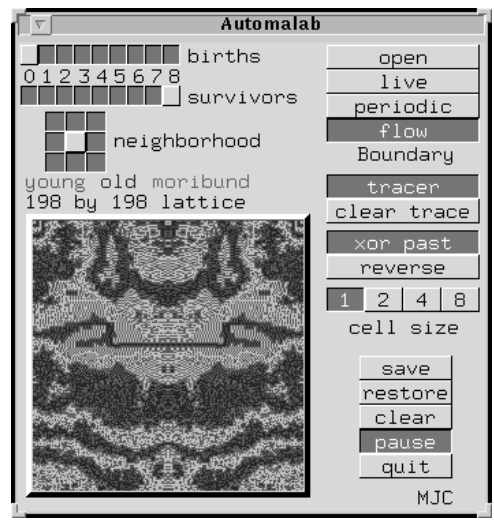

Figure 1. The user interface for xautomalab.

## 2. XISING

Xising illustrates a Monte Carlo simulation of the two dimensional Ising model. This exhibits a second order phase transition from a disordered state at high temperature to an ordered state when cool. The Xwindow display shows the system in the image labeled "spins." This appears above another bit map labeled "changes," representing the spins being changed under the current algorithm. A thermometer indicates the temperature. The known critical temperature is marked on the thermometer. Resizing the window adjusts the lattice dimensions correspondingly.

The program starts with the "local" micro-canonical algorithm of Ref. [1]. A set of "demons" circulates around the lattice trying to flip spins. Each carries a two bit sack of energy ranging from 0 to 16 units in steps of 4. Any energy change associated with a spin flip is compensated by a change in this sack. If the demon's sack cannot accommodate the change, the flip is rejected. The behavior under this algorithm is quite close to that of a conventional Metropolis et al. simulation. The program attains its speed by updating spins one word at a time using multi-spin coding and bit manipulation.

The alternative algorithm constructs a large "cluster" of spins and flips them in unison. This is based on the approach of [2], as adapted to a single cluster by [3]. The particular implementation here is the micro-canonical variation of Ref. [4].

After starting the program, press the heat button and observe how the domains get small and the acceptance, as shown in the "changes" display, gets large. Then press the cool button until the temperature, as displayed in the thermometer, is a couple of tic marks below the critical value. Watch the domains grow as the system magnetizes. Note how the acceptance is largest at the domain boundaries.

At low temperatures a single domain should dominate the system. If, however, bands of different phases wrap around the lattice in either a horizontal or vertical direction, then the system can have difficulty deciding which phase will dominate and can remain meta-stable for a long time. Switching the boundary conditions to antiperiodic forces the system to have at least one domain wall, no matter how cold.

Returning to near the critical temperature, switch to the cluster algorithm. Note how quickly configurations become independent. Heat the system and observe how the typical cluster size becomes quite small. Cooling the system below the critical temperature gives single clusters covering most of the system, which then flashes between dominantly black or white.

To illustrate the power of the cluster algorithm, use the local algorithm to heat the system to a high temperature and then rapidly quench it to somewhat below the critical value. Before the local approach has had time to have the smaller domains dissolve in the dominant one, change to the cluster approach. Note how quickly the cluster sweeps anneal out the included domains.

## 3. XPOTTS

This program illustrates Monte Carlo simulation of the Potts model with $q$ states allowed per site, where $q$ can run from 2 to 255. It is similar to xising except it does not use multi-spin coding. It does add the ability to adjust an applied field using a sack of magnetization carried by the demon. For $q$ larger than 4, the model has a first order phase transition. For large $q$ this is easily observed via a phase separation in the simulations. For more information and suggested experiments, see the text file accompanying the source.

## 4. XAUTOMALAB

Play God over your own universe. With xautomalab you control the micro-physics of a discrete world of cellular automata. This two-dimensional land is an array of colored cells on the screen. Each cell can be alive or empty, with evolution occurring in discrete time steps. Empty cells are grey, newborn ones are red, older living ones are blue, and cells that have just died are green. When the tracer is turned on, old life leaves a legacy by shading the background color.

In one time step, the fate of each cell depends on the number of living neighbors. Using Boolean gadgets, you control when a new live cell will be born on an empty site and when a living one will survive. Another set of buttons determines which neighbors, up to the eight nearest, are included. When the system is paused, a click outside the play field or any specific gadget will update the system a single step. Finally, you can toggle individual cells on and off by pressing a mouse button and sketching over your world.

The save button writes "xautomalab.gif," a standard gif file that you can print or manipulate with any graphics program that likes gif files. The restore gadget will reload a previously stored configuration. The rule used to create a stored configuration is not itself stored or restored. (Any valid GIF87A file can be loaded, but the color information is not used.)

Since sketching when the cells are small is rather imprecise, the save/restore buttons are useful for creating special initial configurations. Using the big block size, sketch your fancy spaceship configuration while the system is paused. Save it, and then switch to a smaller block size. Finally reload the picture at this new resolution.

With an eight cell neighborhood there are 18 birth/survivor buttons. This gives $2^{18} = 262,144$ possible rules. Other neighborhoods give many more, all selectable with the mouse. Indeed, it is unlikely that you will be able to try them all.

In addition, the number of possible universes is further doubled using the "xor past" gadget. When this is activated, the new state is finally XOR'ed with the history one time step back. Thus if a cell was alive in the past, the new state is the opposite of what the birth and survivor gadgets want. The purpose in this is to produce reversible rules. If the history and current states are interchanged, the system will go backwards through the sequence of configurations from which it came. An analogy is reversing all the momenta of a bunch of atoms. This interchange is accomplished by the "reverse" gadget, which interchanges young and moribund cells. If the xor past button is not selected, the reverse gadget does not appear.

If this seems confusing, try this: With the xor past gadget set, clear the screen, draw some small picture, select a random rule from the birth and survivor gadgets, and let the system evolve until the screen becomes a mess. Then hit the reverse button and watch the initial picture reappear. Try repeating this experiment, but alter a single pixel with the mouse at the time of the reversal.

A particularly well known rule is Conway's classic cellular automaton model "life." Here a new cell is born for exactly three neighbors, while a living cell dies with less than 2 (lonely) or more than 4 (overcrowding) neighbors.

Another well-known rule is Fredkin's modulo two model which uses the 4 cell neighborhood. Here a state flips if it has an odd number of active neighbors and is unchanged otherwise. Start with some small picture and observe how the initial state is replicated.

If you want to try a rule running from a random start, run for a while with some chaotic rule (i.e. most rules with births on one neighbor) and then switch to your rule of choice.

Xautomalab is based on my earlier Amiga program Automalab, which appeared on the May 1991 issue of Jumpdisk. That version used direct access the Amiga graphics chips for speed. Comparing this X version to the previous illustrates the awesome power of the Amiga Blitter.

## 5. XSAND

This simulates the sandpile automaton of Bak, Tang, and Wiesenfeld [5]. The model illustrates the concept of self organized criticality. Versions of this program generated the pictures in several popular books [6].

The updating rule is extremely simple. Each cell of a two dimensional lattice contains an integer amount of sand between 0 to 7, inclusive. If this value exceeds 3, that cell is regarded as "unstable" and for the next time step it takes four of its grains of sand and places one on each of its neighbors. The updating is simultaneous for all cells. The total amount of sand is conserved except at the boundary.

The basic idea of self organized criticality is that after lots of random addition of sand followed by relaxation, the system will automatically enter a critical state where the size of an avalanche created by additional sand addition is unpredictable without actually running the process. The sizes of the ensuing avalanches statistically have a power law distribution without any characteristic scale.

The trace buttons allow one to follow the progress of an avalanche. When the tracer is active those sites which "tumble" are flagged and colored "cornflower blue" on attaining stability.

The "double" button doubles all heights modulo 8. This is a convenient way to quickly add lots of sand. The "auto-d" button causes the system to automatically double in height whenever all sites become stable.

The color bar shows the colors representing the various heights. These squares are also gadgets, and clicking on them sets the current pen color. If any mouse button is pressed while over the lattice, you can sketch with this color. To start an avalanche, just select a color larger than 3. The "fill" button sets the entire system to the current pen color.

The save button saves the current configuration as "xsand.gif." The restore gadget will reload a previously stored configuration. The "+ saved" button adds the saved configuration to that presently displayed, modulo 8 on each cell.

With the mouse, scribble some sand randomly on the system. Then repeatedly hit the "double" button to fill the lattice with a random mess. Wait a few minutes until the system stabilizes and activity ceases. Now you should be in the critical ensemble. Make the active color 4, and click the mouse over the lattice. This will start an avalanche, which unpredictably might be large or small. After a few avalanches, turn on the trace button and make some more. Now you can follow where the avalanche has passed.

Note that the avalanche regions always wind up simply connected, with no untumbled islands left over. This is a theorem, and is true for any state in the critical ensemble, but not an arbitrary state. Select height 0 or 1 for the pen, scribble over a small region, and then go back to making some more avalanches. Now it should be easy to make islands, because by removing some sand you have most likely left the critical ensemble.

Select height 0 and clear the system. Then make the boundaries sandy until the system fills up and nothing changes any more. Switch back to open boundaries to let the excess sand run off. Try doubling the final state and letting it relax back. Note that it returns exactly to itself. This state is the unique one in the critical ensemble with this property. Indeed, group theoretically this state represents the identity [7].

Clear the system. Run one or two steps with sandy boundaries, and then go back to open boundaries. Turn the autodouble button on, sit back, and enjoy the show. This also yields the identity state.

Fill the system with height 2. Now draw a picture with height 3. Put the boundaries sandy for a few steps, and then open them up again. Note how the original picture is eventually restored. This is a property of any state in the critical ensemble. Indeed, this is a way to test that a state is critical [8]. Try drawing some more with height 0 or 1. Depending on the picture, the above experiment may or may not mess up your picture.

Save the identity from the earlier experiments. Draw a picture using only heights 2 and 3, as in the previous experiment. Use the "+ saved" button to add in the identity. After a while your picture should magically reappear.

Fill the system with height 2 and go to periodic boundaries. Sketch a while with height 4. With enough sketching, the avalanches will no longer stop. The resulting dynamics can be hypnotic.

## 6. XFIRES

This program simulates forest fires. On each site of a lattice is either nothing, a tree, or a fire. In one time step a fire spreads to adjacent trees and leaves an empty space. Trees are born in a random matter with a probability of approximately 1/32 per time step. If no fires are active, one is started at a random location. Fires can also be started with the mouse button. The Amiga version of this program was published in the Dec. 1993 issue of Jumpdisk.